\documentstyle[aps,prl,twocolumn]{revtex}
\begin{document}
\bibliographystyle{unsrt}

\title{Incremental expansions for Hubbard-Peierls systems} 

\author{ Jiri Malek, Konstantin Kladko and Sergej Flach}

\address{Max-Planck-Institute for Physics of Complex Systems, Noethnitzer Str. 38, 
D-01187 Dresden, Germany 
}

\date{\today}
\maketitle
\begin{abstract}
The ground state energies of infinite half-filled Hubbard-Peierls chains
are investigated  
combining incremental expansion with
exact diagonalization of finite chain segments. 
The ground state energy of equidistant infinite Hubbard ( Heisenberg)
chains  is calculated with a 
relative error of less than $3 \cdot 10^{-3}$  
for all values of $U$ using diagonalizations of $12$-site ($20$-site) chain segments.  
For dimerized chains 
the dimerization order parameter $d$ as a function of the
onsite repulsion interaction $U$ has a maximum at nonzero values of $U$,
if the electron-phonon coupling $g$ is
lower than a critical value $g_c$.
The critical value $g_c$ is found with high accuracy to be $g_c=0.69$.
For smaller values of $g$ the position of the maximum of $d(U)$ is
approximately $3t$, and rapidly tends to zero as $g$ approaches $g_c$
from below. 
We show how our method can be applied to calculate  breathers for the problem of phonon dynamics
in Hubbard-Peierls systems.
\end{abstract}
\pacs{} 

The effect of correlations on the Peierls transition has been one of  
challenging problems in the theory of quasi-one-dimensional compounds.
One of the most important theoretical treatments of the Peierls transition
goes back to the solution of the 
exactly
solvable model of 
noninteracting fermions  proposed by   Su, Schrieffer and Heeger (SSH) \cite{ssh}  
Although being successful in explaining a number of properties of real
quasi-one-dimensional systems, the SSH model
is in a clear disagreement with such experimental results as the emergence
of negative spin magnetization densities for neutral solitons \cite{thomann}. 
One is faced
with a necessity to treat the Coulomb interaction in the electron subsystem.
This interaction should be accounted for
by including a positive Hubbard onsite interaction 
term in the SSH model. We refer to this extended model as to  Peierls-Hubbard model (PHM).
Due to strong one-dimensional quantum fluctuations a mean field theory calculation of PHM
gives qualitatively wrong results, predicting a constant dimerization 
for small $U$ and the abrupt disappearance
of a bond order wave state upon increasing $U$ above a certain threshold
at half-filling \cite{falikov}. Including many-body effects it has been shown by many
authors (\cite{malek} and citations therein) that  
the dimerization $d$ first increases up to  
a maximum 
and then decreases with further increase of $U$. 

It is very difficult to perform an accurate exact diagonalization
investigation of the Peierls transition in the correlated regime .
In the framework of a  standard exact diagonalization approach the required
cluster sizes are found to be far outreaching the capabilities of  modern
computer systems.  Calculations which were performed using available cluster sizes
are drastically depending on the boundary conditions (see for instance \cite{waas}
) and the final conclusions had to be  based on  
extrapolations. The basic questions about the 
value of $g_c$, the behaviour of the system near the critical point, 
the position and the value of the dimerization maximum $U_{max}$ as a function of
the electron phonon coupling remained unanswered. The lack of accurate numerical results
was making it hard to identify the values of model parameters for real systems.    
We show that by combining  an incremental
expansion technique (IET) with numerically exact diagonalizations,  one can overcome the 
abovementioned
difficulties and perform a reliable numerical calculation
of correlated one-dimensional Peierls systems in both strong and weak correlation regimes.   

The quantum chemical method of increments found recently a wide region of application in 
condensed matter (see \cite{fulde1} and references therein). The IET starts  
with splitting a given Hamiltonian operator $H$  into an 
unperturbed part $H_0$  and a number of perturbations $H_1+H_2+...$,
$H = H_0 + \sum H_i$. 
The hierarchy of increments is defined in the following way. 
The first order increment $I^{(1)}_k$ is  given by taking the ground state energy 
${\mathcal E}_k$ of the Hamiltonian $H_0 + H_k$ and subtracting from it the 
ground state energy ${\mathcal E}_0$ of $H_0$, $I^{(1)}_k = {\mathcal E}_k -{\mathcal E}_0$. 
Phenomenologically this increment represents  
the action of the perturbation $H_k$ separately
from all other perturbations $H_i$. 
The total change of the unperturbed ground state energy in first incremental order is then
given by $\sum_i I^{(1)}_i$.
The second order increment $I^{(2)}_{k,l}$ is defined by taking the
ground state energy ${\mathcal E}_{k,l}$ of the Hamiltonian $H_0 + H_k + H_l$ and subtracting from it
${\mathcal E}_0$ {\it and} the first order increments $I^{(1)}_k$ and 
$I^{(1)}_l$, $I^{(2)}_{k,l} = {\mathcal E}_{k,l}-  I^{(1)}_k - I^{(1)}_l - {\mathcal E}_0$.
The increments $I^{(2)}_{k,l}$  
represent the difference between the combined
action of a pair  $H_i$, $H_k$ and the sum of the
uncorrelated actions of both perturbations. In a similar manner  
higher order increments are defined. 
The change of the unperturbed ground state energy of the full system is exactly given by the sum over all
increments (all orders!). Since the increments are usually calculated numerically,
the incremental expansion can be performed up to some given order. 
This expansion is  nonperturbative since the increments
are not related to some small parameter of a perturbation theory. 
The idea of the incremental expansion is similar to Faddeev's treatment of the 3-body problem where 
the unknown 3-body scattering matrix is expressed through the exactly known 2-body scattering matrices.
The discussion of the interrelation of increments and Faddeev equations 
and also the derivation of the incremental
expansion by a resummation of the perturbation theory is given in (\cite{kladko}).

Now we apply the outlined ideas to the PHM Hamiltonian.
This Hamiltonian  is given by a sum of electronic and lattice parts $H = H_{el} +H_{lat}$.
The electronic part in fermionic second quantization form  is given by
\begin{equation}
\label{1}
H_{el} = \sum_{i,\sigma} t_{i} ( c_{i,\sigma}^{\dagger} c_{i+1,\sigma} +h.c.) + 
U\sum_i ( n_{i,\uparrow} n_{i,\downarrow})\;\;.
\end{equation}
Here $U$ is an  onsite Hubbard repulsion matrix element and $t_i$ is the  hopping matrix element  
between the $i$-th and $i+1$-th sites. We consider the case of one electron per site (half filling). 
In the harmonic approximation the lattice part reads as
$H_{lat} = 1/2 K \sum_i v_i^2$.
Here $v_i$ is a bond-length change (see e.g. \cite{malek}) and
$K$ is the spring constant. 
The electron-lattice interaction is assumed to be of the form
$t_i = - (t - \gamma v_i)$.
The strength of the electron-phonon interaction is measured by 
the dimensionless coupling $g = \gamma / \sqrt{K t}$. 
Solving the PHM amounts to finding a minimum of the full energy of the system considered
as a functional of the bond length changes  $v_i$. A remarkable proof of Lieb and Nachtergaele \cite{lieb2} 
tells us that the minimum configuration has to be a dimerized state with alternating 
bond lengths $v_i = (-1)^i v_0$. 
In the following we will use the dimensionless dimerization $d =  v_0 \sqrt{ K / t }$ (see ref. \cite{waas}, \cite{malek}).

Let us now formulate the incremental expansion of the PHM.
A dimerised state represents a sequence of alternating weak and strong 
bonds formed by a modulation of the transfer integral $t_i$. 
It is
natural to cut all the weak bonds and to consider the remaining set of noninteracting 2-site dimers 
as an unperturbed Hamiltonian
$H_0$. The weak bonds are  considered as a perturbation.
The unperturbed Hamiltonian $H_0$ is written as
\begin{equation}
\label{4}
H_0 = \sum_{k = - \infty}^{\infty}  t_{2k} \left( c^{+}_{2k,\sigma}c_{2k +1,\sigma} + h. c.\right) +  
U\sum_{i= - \infty, \sigma }^{\infty} {   n_{i,\uparrow} n_{i,\downarrow}}   
\end{equation}
The ground state of the Hamiltonian operator $H_0$ is known exactly  and is a nondegenerate 
spin-singlet state $S=0$ formed
by a set of noninteracting dimers having two electrons per dimer.
The PHM hamiltonian $H$ is  a sum of $H_0$ and a number of perturbations, formed by the
(weak) bonds  linking neighbouring
dimers 
$H =  H_0 + \sum_{k}{V_k}$, $V_k = \sum_{\sigma} t_{2k-1}(c^{+}_{2k-1,\sigma}c_{2k,\sigma} + h. c.)$.
The incremental expansion is  generated in the following way.
The first order increment corresponds to a bond inserted between two neighbouring 
dimers. Due to general principles outlined above it reads as:
$I^{(1)}= E_4-2 E_2$ ; where $E_{2n}$, $n = 1,2, ... $ denotes the ground state 
energy of a $2n$ sites segment cut out of the infinite chain. 
The second order increment is  defined for a triple of
neighbouring dimers and follows from inserting two bonds into $H_0$. 
To find it one needs to subtract from the energy of three linked dimers the increments
corresponding to 2 pairs of dimers and 3 single dimers in it, hence the expression reads as: 
$I^{(2)}= E_6 - 2I^{(1)} - 3E_2 = E_6 - 2 E_4 + E_2$.
Note that in the incremental expansion only connected clusters of dimers yield  nonzero increments, since
the energy of a disconnected cluster is just the sum of the energies of its parts.
The expressions for  higher order increments are found in similar fashion. 
Due to the choice of $H_0$ (see (\ref{4}) ) the increments do not depend on site indices.
One proves  
by induction that the expression for the n-th order increment, $n>2$ reads as
\begin{equation}
\label{7}
I^{(n)} = E_{2n+2} -2 E_{2n} + E_{2n-2}\;\;. 
\end{equation}
In order to find the ground state energy  of the infinite system one needs to
count the order of increments of each order per dimer. In the
infinite 1-d lattice which we consider, there exists exactly one increment of  each order per
dimer (one makes a one-to-one correspondence between dimers and increments of each
order, assigning each increment to its left most dimer). Therefore, taking into account that
there are two sites per each dimer, the value of the
ground state energy of the infinite lattice per site is written as
${\mathcal E}= 1/2 ( E_2 + \sum_{n=1}^{N} I^{(n)})$.
Here $N$ is the number of increments taken into account.

The feature that the number of increments of any given order per site is constant is an exclusive     
property of one-dimensional lattices. In higher dimensions the per site number of increments of a given order
grows rapidly with the order of the increment.
This special property of one dimension leads to the following result:
\begin{equation}
\label{9}
{\mathcal E}(N) = \frac{1}{2}\left( E_{2 N + 2} - E_{ 2 N}\right) \;\;.
\end{equation}
Formula (\ref{9}) is quite remarkable, since the calculation of ground state
properties amounts to the exact diagonalization of two open chains whose length differs by two.
Note that  expressions of the type (\ref{9}) were  intuitively used before in quantum chemical
calculations (see for instance \cite{dolg}). 

To check our method we first performed a calculation of the ground state 
energy of an equidistant Hubbard infinite
chain at half filling, where the solution is known exactly \cite{lieb}. 
The equidistant case is the worst case for the
method described above, since all the bonds have the same strength.  
The per site value of the  ground state energy ${\mathcal E}$  was calculated using the
formula (\ref{9}) with the incremental order $N=1,2,3,4,5$. The calculation was performed using a Lanczos algorithm.
The results for $N=5$ are shown in the Fig. 1, 
where the exact ${\mathcal E}(U)$ dependence for $t=1$ (solid line)  is plotted 
against the results of (\ref{9})  (open circles).
The relative error $R_E$ decays algebraically with growing order of increments
$R_E=A(U)[2N+2]^{-\nu (U)}$. The exponent $\nu (U)$ and the prefactor $A(U)$ are plotted
in the inset of Fig.1. We find $\nu (U) \geq 2$ for all values of $U$.
Note that $A(0)/A(\infty) \approx 2$ which implies that our results converge faster
for large $U$. Note also that the errors are very small - for $N=6$ typically below
$0.1\%$. This has to be compared with a recent Density Matrix Renormalization Group (DMRG) calculation 
of the same system \cite{dmrg},
where system sizes up to 122 and extrapolations had to used to achieve comparable
precision.

Next we show the results of calculations of 
the dimensionless dimerization  in the PHM as a 
function of $U$ and $g$ (Fig.2).
For $U=0$ the value of $d(g)$ is known exactly \cite{ssh} (see filled symbols at $U=0$ in Fig.2).
An analysis of the relative error $R_d$ of determining $d$ with the help of (\ref{9})
yields exponential convergence  
$R_d\approx {\rm e}^{-\lambda(g) (2N+2)}$. The dependence of $\lambda(g)$ is shown in the
inset of Fig.2. A crossover is detected around $g=0.4$ with $\lambda$ being suppressed to rather
small values for $g \leq 0.4$. That implies that for small values of $U$ the IET method using exact diagonalizations 
is confined to values of the coupling constant $g > 0.4$ if sufficient precision is
requested.
In Fig.2 we present the dependence of $d(U)$ for $g=0.5$, $0.6$ and $0.7$ (open symbols).
For  $g=0.5$, $0.6$ the dimerization $d$ first increases with $U$, and then decreases after
reaching a maximum. For $g = 0.7$ is a monotonously decreasing function of $U$. Therefore
the system has a qualitatively different behaviour for weak and strong couplings $g$, 
as predicted by the GA theory.
On the other hand, our results well agree with the extrapolated values  of
$d$ obtained within the solitonic approach \cite{malek}.

We performed more calculations of $d(U)$  to obtain the dependence
of the position of the maximum $U_{max}$ on the coupling $g$ (see Fig.3). 
Especially we find $U_{max}(g)$ to be a monotonously decreasing function with $U_{max}=0$ at
a critical coupling $g_c$.
Since our method yields very small errors for $g \geq 0.6$ we can estimate the critical
coupling $g$ where $U_{max}=0$ with high accuracy $R_d < 10^{-3}$. It is found $g_c=0.69$. 
The GA prediction $g_c > 0.74$ overestimates this result slightly.
The GA 
gives the position of the dimerization maximum as $U_{max} = 4 t $  for $g$ far below $g_c$.
Our numerical calculation gives $U_{max} \simeq 3 t $.
The small $U$ behaviour of $d$ is found to be $d \sim U^2$, which is consistent with the GA approach.
Furthermore the GA approach predicts that $d$ is an analytic function of $U^2$.
Then it follows that  
$U_{max}(g)$ close to $g_c$ varies as $\kappa (g_c-g)^{1/2}$, which is what we find in Fig.3 ($\kappa \approx 8.25$

For large values of $U$ the system is equivalent to
the Heisenberg spin-exchange model, with $J_i$ given by $J_i = 4 t_i^2 /U$. 
We have calculated the spin-Peierls
transition in this system separately, using the formula (\ref{9}) 
and $N=8$. The results
for the dimerization are plotted  in Fig. 2 (filled symbols). 
Note that the dimerization $d(U)$ for the Hubbard and Heisenberg chains converge for
large $U$ which supports the correctness of our calculations.
For the spin Peierls transition Inagaki and Fukuyama \cite{kuboki} 
found an asymptotic formula 
\begin{equation}
\label{10}
\frac{32 g^2}{\pi \sqrt{1+D}} \left( \frac{t}{U} \right)^{3/2}
\end{equation} 
where $D$ is a constant which was assumed to be of the order of 1/2. 
Our results confirm this choice.
 
The above results show that the IET can be a key method for numerical studying  the static properties
of one-dimensional Peierls systems. 
Our recent calculations show that it can be equally good applied to spin-Peierls systems with
frustration \cite{newpaper}. We believe also that this method could be 
applied to higher dimensional systems, namely
to the Peierls transition in the two-dimensional Hubbard model. 
The cancellation
of the lower order energies does not take place in the two-dimensional Hubbard model 
so the analogue of the
formula (\ref{9}) contains energies of the clusters of all sizes.

To further underline the applicability of the IET, we consider
{\it dynamical} properties of finite Hubbard-Peierls systems.
The dynamics of classical degrees of freedom (phonons) $Q_i$ interacting on a lattice
generically allows for time-periodic and spatially localized solutions namely
discrete breathers (DB) if
the equations of motion incorporate nonlinear terms (see \cite{flach} and citations therein).
These DB solutions can be localized on as few as three neighbouring sites.
If the electron-phonon coupling is taken into account, and the Born-Oppenheimer
approximation is used, the electronic subsystem generates an additional potential
for the classical phonon degrees of freedom. To numerically find again DB solutions,
one needs the electronic energy ${\mathcal E}(\{Q_i\} )$ as a function of the phonon degrees of freedom. 
For a lattice with $L$ sites this amounts to calculating the ground state energy of the electronic
system $L$ times on each time step in order to find the gradient of $H_{el}$.  
Precalculating  the function
$H_{el}$ on a grid is also impossible since it is a 
function of prohibitively many variables. In the static  dimerization case, 
where it is known that the target state is a 
bond conjugate state this problem is avoided since there is
only one variable $d$. 

Again the IET helps to  overcome this problem. Consider a finite chain with periodic
boundary conditions. 
The first order increment $I^{(1)}(x)$, which does not depend on the site index, is obtained 
by fixing all the $Q_i=0$ except one
with $Q_l=x$,
and calculating the change of the electronic energy as a function of $x$, 
$I^{(1)} (x) = {\mathcal E}(x) -{\mathcal E}(0)$. The second order increment is 
obtained by fixing all phonon variables $Q_i=0$  except $Q_l=x$ and $Q_m=y$.
Then the energy of the electronic system will depend on $k = l - m$ , $x$ and $y$. The 
second order increment  reads as $I^{(2)}_k(x,y)  
= {\mathcal E}(x,y) - I^{(1)}(x) - I^{(1)}(y) - {\mathcal E}(0)$. In the same
manner higher order increments are found. 
Our calculations (see \cite{newpaper2} for a detailed discussion) 
show that taking into account increments of first and second
order is enough to calculate the ground state energy of a 14-site Hubbard chain
with periodic boundary conditions for an arbitrary configuration of
$\{ Q_i \}$. The relative error is less than $10^{-3}$.   
The increments are calculated on a two-dimensional  grid to generate
smooth functions. With the help of these 
functions the lattice dynamics can be calculated using
ordinary molecular dynamics techniques. 
 
In this paper we
combined the IET with an exact diagonalization method.
It is known that the 
DMRG is especially accurate when applied to large but finite open chains,
where one can achieve higher and
higher accuracy by iteratively repeating the DMRG procedure \cite{white}.
Taking this into account we think 
that the combination of the IET with the DMRG technique can significantly improve calculations.

We thank Prof. Peter Fulde for  fruitful discussions and continuous support.
\newpage
\noindent
FIGURE CAPTIONS
\\
\\
\\
Fig.1 
\\
\\
Ground state energy of the equidistant Hubbard model
at half filling. Solid line - exact result \cite{lieb}, open
circles - IET result for $N=5$ (see text).\\
Inset: $U$-dependence of prefactor $A$ and exponent $\nu$ of the found
functional dependence of the relative error on $(2N+2)$ (see text).
\\
\\
\\
Fig.2
\\
\\
Dimerization versus $U$ for different values of $g$. Open symbols -
IET results for $N=5$, filled symbols for $U=0$ - exact results,
filled symbols for $U > 5$ - results for Heisenberg chains with
IET and $N=8$. Solid lines are guides to the eye. \\
Inset: $g$-dependence of the exponent $\lambda$ of the found functional
dependence of the relative error on $(2N+2)$ for $U=0$ (see text).
\\
\\
\\
Fig.3
\\
\\
Dependence of $U_{max}$ on the coupling $g$ for the 
Hubbard-Peierls chains with IET and $N=5$.
Circles - numerical results; line - fit (see text).

\end{document}